\documentclass[twoside,english,nofootinbib,preprintnumbers,aps,onecolumn,notitlepage,11pt]{revtex4}
\usepackage{euscript,amssymb}
\usepackage{amsthm}
\usepackage{graphicx}
\usepackage{amsfonts}
\usepackage{amsmath}
\usepackage{amssymb}
\usepackage{fancyhdr}
\usepackage{epsfig}
\usepackage{color}
\usepackage{soul}
\usepackage{calc}
\usepackage{enumerate}
\usepackage{afterpage}
\usepackage[papersize={8.5in,11in}]{geometry}

\allowdisplaybreaks

\newcommand{\be}{\begin{equation}}
\newcommand{\ee}{\end{equation}}

\begin{document}

\title{The Petrov type D equation on genus $>\ 0$ sections of isolated horizons}
\author{Denis Dobkowski-Ry{\l}ko}
	\email{Denis.Dobkowski-Rylko@fuw.edu.pl}
	\affiliation{Faculty of Physics, University of Warsaw, ul. Pasteura 5, 02-093 Warsaw, Poland}
	\author{Wojciech Kami\'nski}
	\email{Wojciech.Kaminski@fuw.edu.pl}
	\affiliation{Faculty of Physics, University of Warsaw, ul. Pasteura 5, 02-093 Warsaw, Poland}

\author{Jerzy Lewandowski}
	\email{Jerzy.Lewandowski@fuw.edu.pl}
	\affiliation{Faculty of Physics, University of Warsaw, ul. Pasteura 5, 02-093 Warsaw, Poland}
\author{Adam Szereszewski}
	\email{Adam.Szereszewski@fuw.edu.pl}
	\affiliation{Faculty of Physics, University of Warsaw, ul. Pasteura 5, 02-093 Warsaw, Poland}
\begin{abstract}   The Petrov type D equation imposed on the $2$-metric tensor and the rotation scalar   
of a cross-section of an isolated horizon can be used to uniquely distinguish the Kerr - (anti) de Sitter spacetime 
in the case the topology of the cross-section is that of a sphere. In the current paper we study that equation on 
closed $2$-dimensional surfaces that have genus $>\ 0$. We derive all the solutions assuming the
embeddability in $4$-dimensional spacetime that satisfies the vacuum Einstein equations with (possibly
$0$) cosmological constant. We prove all of them  have constant Gauss curvature and zero rotation. Consequently, we provide a quazi-local argument for a black hole in 4-dimensional spacetime to have a topologically spherical cross-section.
\end{abstract}

\date{\today}

\pacs{???}

\maketitle
\section{Introduction} A quasilocal local theory generalizes  black holes to surfaces that have some properties
of black hole horizons \cite{Hawking-Ellis,Wald}. They are apparent
horizons \cite{Hayward}, non-expanding horizons, isolated horizons
\cite{LivRevIH,Lih,ABL1, ABL2, LPhigh,KLPhigh}, Killing horizons
\cite{Wald,rw}. Typically, the generalized black holes have infinite
set of local degrees of freedom as opposed to the finite dimensional
families of the Kerr, Kerr-de Sitter, and Kerr-anti de Sitter spacetimes \cite{Kerr,dem}. For example, the  internal geometry of every vacuum non-extremal isolated horizon $H$ is determined by a metric tensor $g$ and a rotation $1$-form  potential $\omega$ induced on a spacelike section $S\subset H$ \cite{ABL1}. They are unconstrained, a priori free (in the extremal case that changes drastically \cite{LPextremal}).  On that data  we imposed  "the Petrov type D equation" implied by assuming that the spacetime Weyl 
tensor is of the Petrov type D at the horizon, and that it is Lie dragged by the null symmetry of the horizon geometry  \cite{LPkerr,DLP1}. 
The equation was solved explicitly in the case of $S$ topologically equivalent to two-dimensional sphere $S_2$ and axially 
symmetric data $(g,\omega)$ \cite{LPkerr,DLP2}. The corresponding isolated horizons are embeddable isometrically
in  the Kerr, Kerr-de Sitter, or Kerr-anti de Sitter spacetimes \cite{dem} depending on the cosmological constant, or in the near extremal
Killing horizon limit spacetimes known under the name Near Horizon Geometry \cite{Horowitz,PLJ,LivRevNHG,Podolsky1,Podolsky3,AszerLewWal2}.  If the isolated horizon is bifurcated  and both the components are of the Petrov type D, then the geometry of the horizon is necessarily axially 
symmetric \cite{Raczaxial,LSaxial}. That is proven locally, without the rigidity theorem.  In that way our results become part of the wider context of local characterizations that can be used to distinguish  those globally defined spacetimes  
\cite{Mars1,Mars2,coley1,coley2,coley3}.

The Petrov type D equation also turns out to be a necessary  integrability condition for the Near Horizon Geometry equation \cite{DLP1}.  
              
In the current  paper, we consider the Petrov type D equation for the case when  $S$ is a two-dimensional closed surface 
of the genus $>\ 0$. We derive a general solution for arbitrary value of the cosmological constant.

\section{ The Petrov type D equation} 
The Petrov type D equation is imposed on  a Riemaniann metric tensor $g_{AB}$ and  a $1$-form $\omega_A$
defined on a two-dimensional manifold $S$.  
The equation involves  two scalar invariants of that data: the
Gaussian curvature $K$ of the metric $g_{AB}$ and the pseudo-scalar
$\Omega$ of $\omega$, 
\begin{equation}
R_{AB}\ =:\ Kg_{AB}, \ \ \ \ \ \ \ \ \ \ \ \ d\omega\ =:\ \Omega \eta,
\end{equation} 
where $R_{AB}$ and $\eta$ are the Ricci tensor and the area 2-form, respectively, of  the metric tensor $g_{AB}$. 
The equation  uses  the complex structure defined by $g_{AB}$, hence it is convenient to express it by a  complex  null 
co-frame $m_A$, such that  
\begin{equation}
g_{AB} = m_A \bar m_B + m_B \bar m_A ,\ \ \ \ \ \ \ \ \ \ \ \ \eta_{AB} = i(\bar m_A m_B - \bar m_B m_A).
\end{equation}
With that notation,  and with the  covariant derivative $D_A$ defined by $g_{AB}$  such that,
$$ D_Ag_{BC}\ =\ 0,  \ \ \ \ \ \ \ \ \ \ \ \ (D_AD_B - D_BD_A)f\ =\ 0$$
for every function $f$, the Petrov type D equation  reads \cite{DLP1}
\begin{equation}\label{typeD}
\bar m^A \bar m^B D_A D_B \bigg ( K - \frac{\Lambda}{3} + i \Omega \bigg)^{-\frac{1}{3}} = 0,
\end{equation}
where $\Lambda$ is a constant. We are also assuming the non-degeneracy condition 
\begin{equation}   \label{condition}
K - \frac{\Lambda}{3} + i \Omega \neq 0
\end{equation}
at every point of $S$. If condition (\ref{condition}) is not satisfied at some point on $S$ then at the corresponding points of the horizon $H$ (see below) the Weyl tensor is either of the Petrov type $N$ or $O$  \cite{DLP1}.

The issue of existence and uniqueness of the cubic root requires some extra care.  For every point $x$, in a contractable open neighborhood there exists a function $f$, as continuous and of the same differentiability class  as $K + i \Omega$, such that 
\begin{equation}\label{f}
f^3\ =\  K - \frac{\Lambda}{3} + i \Omega . 
\end{equation}
It is defined up to a constant factor 
$$f\mapsto e^{\frac{2\pi m}{3}i}f,$$
where $m\in\mathbb{Z}$.  Hence, the equation (\ref{typeD}) does define a global on $S$ unique condition on 
the function $K - \frac{\Lambda}{3} + i \Omega $.  If $S$ is topologically a $2$-sphere, then the function $f$ 
can be  defined globally on $S$, up to the overall constant factor. In this paper we will be considering the type D equation
on an orientable, connected closed manifold $S$ characterized by genus $>0$.   If a given metric $g$ and $1$-form $\omega$
define the function  $K - \frac{\Lambda}{3} + i \Omega$  that does not admit a cubic root continuous on entire $S$,
then, there exists: a $3$-fold covering $2$-manifold  $\tilde{S}$, with the metric $\tilde{g}$ and $1$-form  $\tilde{\omega}$ being pull backs of $g$ and $\omega$ from $S$ such that the corresponding
function $\tilde{K} - \frac{\Lambda}{3} + i \tilde{\Omega} $ does admit a globally defined continuous cubic root on $\tilde{S}$ (different points of a fiber of the covering correspond to different choices of the root on $S$). 
The extension $(S,g,\omega)\rightarrow (\tilde{S},\tilde{g},\tilde{\omega})$ maps solutions to the equation (\ref{typeD})  into solutions
of that equation. The covering manifold is connected, orientable, closed and its genus satisfies $\tilde{g}=3g-2$ (by Gauss-Bonnet theorem) and thus it is bigger or equal the genus of $S$.
Specifically, if $S$ has the topology of  torus, so does $\tilde S$.

The equation is invariant with respect to the gauge transformations
\begin{equation} \label{gauge}
\omega \mapsto \omega+ dh,  \ \ \ \ \ \ \ \ \ \ \ \ h\in C^n(S)  
\end{equation}
and with respect to every diffeomorphism $\Phi:S\rightarrow S$,
\begin{equation} \label{diffinvariance}
(g,\omega) \mapsto (\Phi^*g, \Phi^*\omega).
\end{equation}

A two-dimensional surface $S$ equipped with  a metric tensor $g_{AB}$ and a $1$-form $\omega_A$ determines 
the spacetime geometry at a non-extremal stationary to the second order null  surface $H$ diffeomorphic to 
$S\times \mathbb{R}$ and contained in a $4$-dimensional  spacetime $M$ endowed with a metric tensor $g_{\mu \nu}$ that satisfies the vacuum Einstein equations with 
a cosmological constant $\Lambda$:
\begin{equation}\label{Einstein}
R_{\mu \nu} -\frac{1}{2}g_{\mu\nu}R + \Lambda g_{\mu \nu} = 0.
\end{equation}
By stationary to the second order we mean that there exists a vector field $\ell$ in a neighborhood of $H$ tangent to and null at $H$,  that Lie drags along $H$ the metric tensor $g_{\mu \nu}$, the spacetime covariant derivative $\nabla_\mu$ and the Riemann tensor 
${R_{\mu\nu\alpha\beta}}$:
\begin{equation}\label{stationarity}
{\cal L}_\ell g_{\mu \nu \mid_H} = [{\cal L}_\ell , \nabla_\nu ]_{\mid_H} = {\cal L}_\ell R_{\mu \nu \alpha \beta \mid_H} = 0.
\end{equation}
Since the null generators of a null surface are geodesic curves and the vector field $\ell$ is tangent to the null generators of $H$, $\ell$ is self parallel at $H$ \cite{Wald}:
\begin{align}
\ell^\mu \nabla_\mu \ell^\nu_{\mid_H} = \kappa \ell^\nu,
\end{align}
and by the zeroth law of non-expanding null surfaces thermodynamics \cite{ABL1}, the function $\kappa$
(surface gravity) is constant
\begin{align}
\kappa = \text{const}.
\end{align} 
By non-extremal we mean that
\begin{align}\label{nonextremal}
\kappa \neq 0. 
\end{align}
In this construction the $2$-surface $S$ is a section of $H$ transversal to the vector field $\ell$. The metric $g_{AB}$ is induced 
in $S$ by the spacetime metric $g_{\mu\nu}$.  The $1$-form $\omega_A$  is defined by the spacetime covariant derivative 
of $\ell$ in the directions tangent to $S$, namely, for every vector $X$ tangent to $S$, 
\begin{equation}
X^A\nabla_A\ell^\mu \ =\ X^A\omega_A \ell^\mu .
\end{equation}
By the stationarity assumption (\ref{stationarity}) and the Einstein equations (\ref{Einstein}), $g_{AB}$ and $\omega_A$  induced on $S$ determine at $H$:  $g_{\mu\nu}$, $\nabla_\mu$, and $R_{\mu\nu\alpha\beta}$.   If $S'$ is another section of $H$,
then 
\begin{equation}\label{gauge'} g'_{AB}\ =\ g_{AB}, \ \ \ \ \ \ \ \ \ \ \ \ \omega'_A\ =\ \omega_A + \kappa D_A f \end{equation}
where we identify $S$ and $S'$ by using the null geodesics in $H$, and  $f$ is a function on $S'$.  
If  we hold $H$, $\ell$ and $S\subset H$ fixed, and vary the spacetime metrics  $g_{\mu\nu}$  such that
(\ref{Einstein}, \ref{stationarity}, \ref{nonextremal}) then the data $(g_{AB}, \omega_A)$ ranges all possible 
metric tensors and $1$-forms. 

In the sense explained above, the data $(g_{AB},\omega_A)$ is free on a section $S$ of $H$. In particular
it determines the spacetime Weyl tensor $C_{\mu\nu\alpha\beta}$ at $H$. Now, the Weyl tensor $C_{\mu\nu\alpha\beta}$ at 
$H$ is of the Petrov type D if and only if $g_{AB}$ and $\omega_A$ satisfy the Petrov type D equation (\ref{typeD}).

\section{The Petrov type D equation on a two-dimensional torus}
In this section the $2$-manifold  $S$ a is a two-dimensional torus $T_2$,
\begin{equation}
S\ =\ T_2\ =\ S_1 \times S_1 ,
\end{equation}
where $S_1$ is a circle. We fix on the first copy of $S_1$ a coordinate $\phi\in [0,2\pi )$ and on the second copy a coordinate
$\psi\in [0,2\pi)$. They set coordinates 
$$(x^A)=(\phi,\psi)$$ 
on $S$.  The coordinates are  defined globally on $S$, except for that they are
not continuous at $\phi,\psi=0$. However, the cotangent frame $d\phi,d\psi$ and the dual tangent 
frame $\partial_\phi,\partial_\psi$ 
defined globally on $S$, are continuous and smooth everywhere. 
That property will be  important below.  Modulo the diffeomorphisms (\ref{diffinvariance}), every flat metric tensor $g^{\rm flat}_{AB}$ on $S$ can be written 
in the following form,  
\begin{equation*}
g^{\rm flat}_{AB}dx^Adx^B\ =\  \frac{1}{P^2_0}\left(a^2d\phi^2 + 2ab\, d\phi d\psi + (1+b^2)d\psi^2\right), 
\end{equation*}
where $a,b,P_0$ are  real constants,  $a,P_0\not=0$
\cite{string}.
Every general metric tensor $g_{AB}$ on $S$ is conformally equivalent to a flat one \cite{Weyl}, 
hence,  there exists a non-vanishing function $P$ on $S$ such that
\begin{equation}
g_{AB}dx^Adx^B\ =\ \frac{1}{P^2}\left(a^2d\phi^2 + 2ab\, d\phi d\psi + (1+b^2)d\psi^2\right). 
\end{equation}
For the compatibility with the Petrov type D equation (\ref{typeD}) we assume that $g_{AB}$
and in the consequence  $P$ are at least $4$ times differentiable.
  

To introduce the complex null basis we define complex coordinates $(z,\bar{z})$, 
\begin{equation}
z\ = \frac{a\phi + b \psi +i\psi}{\sqrt{2}}.
\end{equation}
Since they are just linear in $\phi$ and $\psi$, they are defined globally on $S$ with the non-continuity
of the same type as $\phi$ and $\psi$.
 It is easy to check by inspection, that in  the new coordinates, the metric tensor $g_{AB}$ reads
\begin{equation}
g_{AB}dx^Adx^B\ =\ \frac{2}{P^2} dz\,d\bar{z},
\end{equation}
whereas the area $2$-form is
\begin{equation}
\eta = i \frac{1}{P^2} dz \wedge d \bar z.
\end{equation}
The null tangent and co-tangent frame respectively is
$$ m^A\partial_A\ =\ P\partial_z, \ \ \ \ \ \ \ {m}_Adx^A \ =\ \frac{1}{P}d{\bar{z}}.$$
Notice, that despite of the discontinuity of the coordinates $(z,\bar{z})$, the tangent frame 
$(m^A,\bar{m}^A)$, the cotangent frame $({\bar m}_A, m_A)$,  the complex valued vector fields $\partial_{z},\partial_{\bar z}$ 
and the complex valued $1$-forms $dz, d\bar{z}$ are all globally defined on $S$.

In those coordinates the differential operator  featuring in (\ref{typeD}) is a composition
(denoted by "$\circ$") of three operators
\begin{equation}
\bar{m}^A\bar{m}^BD_AD_B\ =\ \partial_{\bar z} \circ P^2 \circ \partial_{\bar z}.
\end{equation}
Hence, the type D equation  (\ref{typeD}) may be  written now in the following explicit form:
\begin{equation}\label{typeD2}
\partial_{\bar z} \big(P^2 \partial_{\bar z} f \big ) = 0,
\end{equation}
where
\begin{equation}\label{f}
f = \left(K - \frac{\Lambda}{3} + i\Omega \right) ^{-\frac{1}{3}}.
\end{equation}
We know from the previous section, that to ensure the continuity of the cubic root, we may need
 to consider a suitable extension $(\tilde{S},\tilde{g}, \tilde{\omega})$. Since topologically, $\tilde{S}$ is the torus again,
 we just drop the tildes  and continue.

The first consequence of the eq. (\ref{typeD2}) and of the global properties of the function $P$ and the vector field 
$\partial_{\bar z}$ is that $P^2 \partial_{\bar z} f$ is an entire holomorphic function on all of $S$. Hence, due the 
compactness  of $S$,  
\begin{equation}\label{holvectorus}
 P^2 \partial_{\bar z} f = F_0 = \text{const},
 \end{equation}
equivalently
\begin{equation}\label{partiazbarf}
\partial_{\bar z} f = \frac{F_0}{P^2}.
\end{equation}
Next, we  perform the following calculation
\begin{equation}
F_0 \int_{S}\eta\ =\ i\int_{S}\partial_{\bar z} f dz\wedge d\bar{z}\ =\ -i \int_{S} d\left(fdz\right)\ =\ 0,
 \end{equation}
where again, the last equation follows  from the compactness of $S$. That implies 
\begin{equation}
F_0 = 0.
\end{equation}
Furthermore, going back to the eq. (\ref{partiazbarf}) we conclude that $f$ is an entire holomorphic
function on $S$, hence it has to be constant,
$$f\ =\ {\rm const}.$$
That is a general solution of the eq. (\ref{typeD2}). 
Recalling, that in our case $f$ is given by the eq. (\ref{f}), we can see that also
\begin{align}
K= K_0 = \text{const}, && \Omega =\Omega_0 = \text{const}.
\end{align} 
But since $S$ is a torus, the Gauss-Bonnet theorem implies about the Gauss curvature 
$$ K_0\int_{S}\eta\ =\ \int_{S}K \eta\ =\ 0,$$
meaning 
$$K_0=0.$$
Similar integral law applies to the rotation scalar $\Omega$, namely
$$ \Omega_0\int_{S}\eta\ =\ \int_{S}\Omega \eta\ =\ \int_{S}d\omega\ =\ 0.$$

Now, we  recover the tildes dropped at the the beginning of that section.  
The conclusion is, that, if $\Lambda\neq 0$, then the extension  $(\tilde{g}_{AB}, \tilde{\omega}_A)$ to the Petrov type D equation (\ref{typeD}) on
$${S}\ =\ T_2$$
is a flat metric tensor $\tilde{g}_{AB}$ and a closed $1$-form $\tilde{\omega}$. But that implies, that the original
$g_{AB}$ and $\omega_S$ on the original  $S=T_2$ are also flat and closed, respectively. Finally, in that case, 
the extension turned out to be trivial.
In the case $\Lambda = 0$ there are no solutions because of (\ref{condition}).


From the point of view of the reconstruction 
of a stationary to the second order null surface from $(g,\omega)$,  the $1$-form $\omega$ is meaningful only
modulo the gauge transformations (\ref{gauge'}). A $2$-torus, however,  admits non-trivial de Rham cohomology 
group. Modulo the diffeomorphisms (\ref{diffinvariance}) and the gauge transformations (\ref{gauge}), a general
solution has the form
\begin{equation}\label{solutions}
g_{AB}dx^Adx^B\ =\ \frac{1}{P_0^2}\left(a^2d\phi^2 + 2ab\, d\phi d\psi + (1+b^2)d\psi^2\right),\ \ \ \ \ \ \omega\ =\ 
Ad\phi\ +\ Bd\psi  
\end{equation}
where $P_0>0, a >0, b, A,B$ are  arbitrary constants.

\section{The Petrov type D equation on higher genus surfaces}

In this section   $S$ has genus $>1$, it is
orientable and closed.  Again, it is endowed with a $2$ metric tensor $g$, and a $1$-form $\omega$. If the function $f$
(\ref{f}) cannot be continuous on $S$, then we consider the suitable extension $\tilde{S}$ and drop the tildes. 
The  extended  solution  of (\ref{typeD}) obtained below, will also determine the extension  as trivial.


We can cover $S$ with charts, such that  in each of them complex coordinates
\begin{equation}\label{gzz}
g_{AB}dx^Adx^B\ =\ \frac{2}{P^2} dz\,d\bar{z},
\end{equation}
and the equation   
\begin{equation}\label{typeD1/2}
\bar m^A \bar m^B D_A D_B f\ =\ 0,
\end{equation}
for arbitrary unknown function $f$ takes the form
\begin{equation}\label{typeD3}
\partial_{\bar z} \big(P^2 \partial_{\bar z} f \big ) = 0.
\end{equation}
The entry of the parentheses can be considered as a component of the
complex vector field 
\begin{equation}   
(P^2 \partial_{\bar z} f) \partial_{z} \ =
  (g^{zA}\partial_Af)\partial_z. 
\end{equation}      
The point is, that this vector field is globally defined on $S$. It is constructed
from the gradient of $f$
\begin{equation}   
(g^{AB}\partial_Af)\partial_B\ =\ (P^2 \partial_{\bar z} f) \partial_{z} + (P^2 \partial_{z} f) \partial_{\bar z}
\end{equation}      
and by a globally defined decomposition of complexified tangent space at each point
\begin{equation} \label{decomp}Y=Y^z\partial_z + Y^{\bar z}\partial_{\bar z} \ \mapsto\ Y^z\partial_z . 
\end{equation} 
The decomposition uses the locally defined  coordinates $(z,\bar{z})$. However,  
a most general  coordinate transformation   that preserves the form of the metric tensor in (\ref{gzz}) 
satisfies
 \begin{equation}\label{trans} \frac{\partial {z'}}{\partial \bar z}=0, \end{equation} 
and therefore preserves the decomposition (\ref{decomp}).

Another operation invariant with respect to the holomorphic coordinate transformations (\ref{trans}) is the
anti-holomorphic derivative $\bar{\partial}$ that acts in the vector space of all the vector fields 
$$X=X^z\partial_z$$ 
with arbitrary point dependent coefficients $X^z$ as follows
\begin{equation}\label{holomorphicvectors}\bar{\partial}
  X\ =\ \partial_{\bar z}X^z\partial_z\otimes d{\bar z}.
 \end{equation} 
 
Solutions $X^z\partial_z$ to the equation 
$$\bar{\partial}X\ =\ 0$$
are called holomorphic vector fields.  For every compact orientable $2$-surface $S$ the dimension of the space  
of the holomorphic vector fields is known \cite{GF}. 
In particular, if the genus $>1$ the dimension is $0$,
$$X=0.$$
Hence,   the eq. (\ref{typeD3}) implies
\begin{equation}   \partial_{\bar z} f\ =\ 0.    \end{equation}  
Then due to the compactness of $S$,
\begin{equation}   f\ =\ {\rm const}  . \end{equation}  
Hence, as in the case of torus, via (\ref{f}) it implies
\begin{equation}
K\ =\ {\rm const}, \ \ \ \ \ \ \ \ \Omega\ =\ {\rm const}.
\end{equation} 
Since on every compact $S$ the rotation scalar  $\Omega$ satisfies the constraint
$$\int_S \Omega\ =\ 0$$
as before we conclude
\begin{equation}
\Omega\ =\ 0. 
\end{equation}
On the other hand $K$ is proportional  to the inverse of the area  ${A}$ with a coefficient given by the Gauss-Bonnet theorem
\begin{equation}   K\ =\ \frac{4\pi(1-{\rm genus})}{A} 
\end{equation} 
except for
\begin{equation}
K=\frac{\Lambda}{3}
\end{equation}
which violates the condition (\ref{condition}).

\noindent{\bf Remark}. The argument about the holomorphic vector could have been used also 
in the torus case: every torus admits a $1$-complex-dimensional space of  holomorphic vector fields.
That observation immediately implies eq. (\ref{holvectorus}). Then, however,  one still has to show
that the constant $F_0$ actually is $0$ in the case of  a  vector field $ (g^{zA}\partial_Af)\partial_z$,
as we showed in Sec. III.

\section{Summary and Discussion}  We derived all the  metric tensors $g$ and  $1$-forms $\omega$
defined on a  compact, orientable $2$-surface of genus $\ge 1$ that are solutions to the Petrov type D equation (\ref{typeD}). 
\medskip

\noindent{\bf Theorem 1}  {\it A pair $(g,\omega)$ is a solution to the Petrov type D equation with  a cosmological constant $\Lambda$
 on a  compact, orientable $2$-surface of genus $\ge 1$ if and only if
$g$ has constant Gauss curvature (Ricci scalar)
$$K\ =\ {\rm const} \neq \frac{\Lambda}{3} $$
and  $\omega$ is closed
$$d\omega\ =\ 0.$$}

\medskip

The assumption about the type D could be relaxed to possible
type O in some degenerate subsets of $S$. Then more solutions
can possibly exist.

The solutions were easy to guess therefore the main result is that there are  no other solutions.  
Still, the family of the solutions is more than zero dimensional. 
For example, for $S=T_2$ the family of solutions is $5$ dimensional. 
The corresponding isolated horizons (stationary to the second order) are non-rotating - 
their angular momentum $J=0$. Therefore, one may conclude, that 
\medskip

\noindent{\bf Theorem 2}  {\it  Every rotating Petrov type D isolated horizon stationary to the second order
and contained in a $4$-dimensional spacetime that satisfies the vacuum Einstein equations
with possibly non-zero cosmological constant, has spacelike section of the topology of a $2$-sphere.}
\medskip

In other words, for spacelike sections of genus $>0$ there are no rotating Petrov type D isolated horizons stationary to the second order and contained in a $4$-dimensional spacetime satisfying the vacuum Einstein equations with possibly non-zero cosmological constant. Due to the vanishing of the rotation scalar $\Omega$,  each of the solutions  satisfies also
the conjugate Petrov type D equation
$$ m^Am^BD_AD_B\left(K-\frac{\Lambda}{3}+i\Omega\right)^{-\frac{1}{3}}\ =\ 0.$$          
Therefore, by the black hole holograph  technique \cite{Racz1,Racz2,LRS} one can construct from each $S$ and $(g,\omega)$ a spacetime
$$M=S\times\mathbb{R}\times\mathbb{R}$$ 
that contains a bifurcated horizon of the bifurcation surface $S$  and such that the Petrov type of the spacetime Weyl tensor is D
at the horizon \cite{LSaxial}.  

The Petrov type D equation is a necessary integrability condition for the Near
 Horizon Geometry equation \cite{ABL1,LPextremal,LivRevNHG,DLP1}

 $$\nabla_{(A}\omega_{B)} + \omega_A\omega_B - \frac{1}{2}K g_{AB} + \frac{1}{2}\Lambda g_{AB}=0$$
 valid wherever 
 \begin{equation}\label{Psi2not0}
 \Psi_2 := K-\frac{\Lambda}{3} + i\Omega\not= 0.
 \end{equation}

  Therefore, all the solutions to the NHG
 equation on the $2$-surfaces of genus $>0$   such that (\ref{Psi2not0}) at every point of $S$ belong to the solutions 
 described by Theorem 1. 
Those observations make a new result:
 \medskip

 \noindent{\bf Corollary 1}  {\it If $(g_{AB}, \omega_A)$ is a solution to the near horizon geometry equation  on a 
 connected, orientable, compact $2$-manifold $S$ of genus$>0$, such that (\ref{Psi2not0}) everywhere on $S$,   then it is  static, and the Gauss curvature is constant, 
 that is
$$d\omega=0, \ \ \ \ \ \ \  K=\rm const.$$}
\medskip
 
That conclusion can be combined  with a body of prior work on the Near Horizon equation for higher genus S. 
To begin with, if 
$$\Lambda \geq 0$$ 
then by the topological constraint  derived in \cite{PLJ}  the only genus $>0$ compact solution is the trivial
solution with $\Lambda=0$, $S=T^2$ with flat metric and rotation form
$\omega_a=0$  (see also \cite{J, LivRevNHG2}  and Theorem 3.1 in \cite{LivRevNHG}).
Hence, in the non-negative $\Lambda$ case the issue of the equation on the higher genus surfaces is  solved in the literature which means our new integrability condition is not needed.

The negative cosmological constant
$$\Lambda<0$$  weakens the topological constraint, therefore
that is the case where one may expect solutions of the genus$>0$. 
However, the axially symmetric solutions  with $S=T^2$ have
been excluded, proven in \cite{Li}. 
Moreover,  static near horizon geometries with any compact $S$ 
must have 
$$\omega_A=0$$ 
and the Gauss curvature 
$$K= \rm const,$$ proven 
in \cite{CRT}.

Our very Corollary 1  combined with the latter result provides  a general  solution to the 
Near Horizon Geometry equation on surfaces with genus$>0$ such that (\ref{Psi2not0}):
\medskip

 \noindent{\bf Corollary 2}  {\it The only solutions  $(g_{AB}, \omega_A)$  to the near horizon geometry equation  on a 
 connected, orientable, compact $2$-manifold $S$ of genus$>0$ such that (\ref{Psi2not0}), satisfy
 $$\omega_A=0, \ \ \ \ \ \ \ K=\ \frac{4\pi(1-{\rm genus})}{A} = \Lambda,$$
 where $A$ is the area of $S$.}
\medskip

Finally,  it is known in the literature  that for every solution to the NHG equation on $S=S_2$, 
the function $\Psi_2$  is either identically or nowhere  $0$ 
\cite{LPextremal,ChruscielTodSzybka} (the argument used there for $\Lambda=0$ easily generalizes  
to the $\Lambda\not=0$ case). With somewhat more work one can show, that $\Psi_2$  is either identically 
or nowhere  $0$ on any compact orientable, $2$-dimensional $S$ \cite{DKLS2}. 

That seems to complete the problem of the NHG equation on genus$>0$ surfaces.

\medskip
 
\noindent{\bf Acknowledgements:}
This work was partially supported by the Polish National Science Centre grant No. 2015/17/B/ST2/02871.

\end{document}